\documentclass{iopart}
\usepackage{graphicx}
\usepackage{iopams}
\begin{document}
\title{A simple approach to the Landau-Zener formula}

\author{Amar C Vutha}
\address{Department of Physics, Yale University, New Haven, CT 06520, USA}
\ead{amar.vutha@yale.edu}

\begin{abstract}
The Landau-Zener formula provides the probability of non-adiabatic transitions occuring when two energy levels are swept through an avoided crossing. The formula is derived here in a simple calculation that emphasizes the physics responsible for non-adiabatic population transfer.
\end{abstract}

\pacs{03.65.Ca, 
      31.50.Gh 	
      }

~\\ 

Energy level crossings have been of continued interest in physics, beginning with early analyses of the Born-Oppenheimer approximation \cite{Zen32} upto present day applications to the production of ultracold molecules \cite{LSB+08} and the understanding of quantum phase transitions \cite{ZDZ05}. The probability of non-adiabatic population transfer during passage through an avoided level crossing is given by the well-known Landau-Zener formula (LZF) \cite{LL03}. The usual derivations of the LZF involve the use of mathematical techniques such as contour integrals \cite{LL03,Wit05} or Weber functions \cite{Zen32}, which might present a barrier to introducing non-adiabatic transitions in a basic quantum mechanics course. However the LZF can be understood using simple undergraduate-level physics, without getting mired in complicated mathematics. In the hope that it may be pedagogically useful and provide some physical insight into non-adiabatic population transfer, here is an elementary derivation of the LZF.

Let $|0\rangle$ and $|1\rangle$ be two basis states of a system. The state vector is $|\psi\rangle = \psi_0 |0\rangle + \psi_1 |1\rangle$. Assume that, at the initial time $t = t_i$, the system is prepared in the $|0\rangle$ state: $\psi_0(t_i) = 1, \psi_1(t_i) = 0$. Let the excited state $|1\rangle$ be higher in energy compared to $|0\rangle$ by an amount $\omega_0$ (we set $\hbar=1$ everywhere for simplicity). Let these two states be coupled by a general time-dependent interaction. In the rotating wave approximation, the Hamiltonian for this two state system can be written as 
\begin{equation} \label{eq:Hint}
H = \left( \begin{array}{cc} 0 & \Omega^{\dagger} e^{+i \omega t} \\ \Omega \ e^{-i \omega t} & \omega_0 \end{array} \right).
\end{equation}
Here $\Omega$ is the strength of the time-dependent interaction, and $\omega$ its frequency.

\begin{figure}[h!]
\centering
\includegraphics[width=0.65\columnwidth]{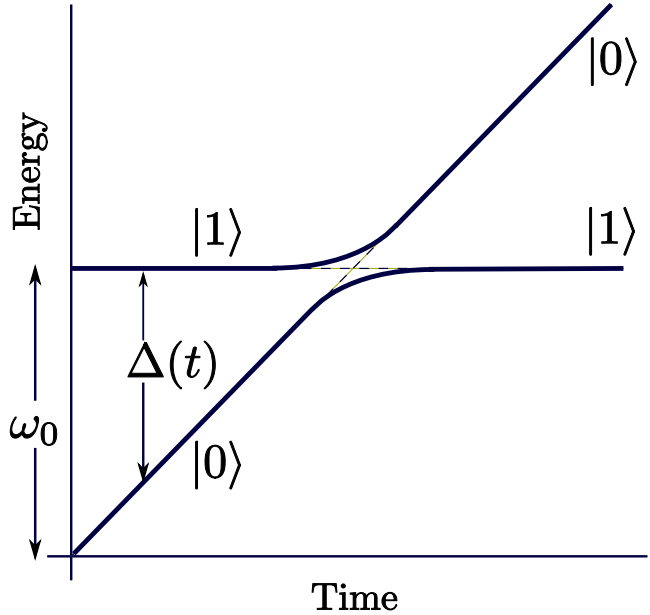}
\caption{Energies of the dressed $|0\rangle, |1\rangle$ states in the rotating wave approximation. When the detuning is swept linearly through resonance, the probability of jumps across the avoided crossing is given by the Landau-Zener formula.} 
\label{fig:one}
\end{figure}

Following Zener \cite{Zen32}, consider a process where the detuning $\Delta = \omega - \omega_0$ is varied linearly with time: $\frac{d}{d t}\Delta(t) \equiv \dot{\Delta} = \mathrm{constant}$. This can be accomplished by varying the frequency $\omega$ of the perturbation $\Omega \ e^{-i \omega t}$, or by tuning the energy level splitting $\omega_0$ (using a magnetic field, for example). For a constant detuning $\Delta$, the dynamics of the system can be completely described by the usual Rabi flopping formula for a 2-state system \cite{BKD08}, but the Rabi formula is not a solution of the Schrodinger equation when the parameters $\omega_0, \omega$ are themselves functions of time. The exact analytic solution when $\Delta$ is linearly ramped in time involves Weber functions \cite{Zen32}. However, we shall find that the physics of this process can be understood without recourse to the mathematics of special functions.

The main simplification arises because a time-varying detuning leads to rapid dephasing of coherent Rabi oscillations. Say we model the linear ramp as follows: $\omega$ varies discretely in very small steps, each of which lasts for some small duration $\delta t$. Within each of these time segments of duration $\delta t$ the detuning is a constant and the coherent evolution of the system is described by the usual Rabi flopping amplitudes. To obtain the state vector of the system for longer durations, the amplitudes from consecutive time-segments have to be added together. However the different detunings, and therefore different Rabi frequencies, in each segment mean that these amplitudes get added together with lots of different phases. Due to this dephasing introduced by the detuning ramp, the population dynamics of the system can be described quite well by summing the transition \emph{probabilities} per unit time instead of the amplitudes. We estimate the dephasing time $\tau_D$ as the duration to accumulate a phase difference of $\sim 2 \pi$ rad between Rabi oscillations in consecutive segments: $\tau_D \sim \sqrt{\frac{4 \pi}{\dot{\Delta}}}$. We will find that exact knowledge of this dephasing time is not necessary to get to the LZF.

Under the action of the time-dependent Hamiltonian $H$, the probability per unit time for excitation out of the $|0\rangle$ state is the transition rate $\Gamma$, given by the familiar expression \cite{BKD08}
\begin{equation}
\Gamma = \Omega^2 \frac{\gamma}{\Delta^2 + \gamma^2/4}.
\end{equation}
Here $\gamma$ is the decay rate of the Rabi oscillations, which we set equal to the inverse dephasing time: $\gamma = 1/\tau_D \sim \sqrt{\frac{\dot{\Delta}}{4 \pi}}$. 

\begin{figure}[h!]
\centering
\includegraphics[width=0.65\columnwidth]{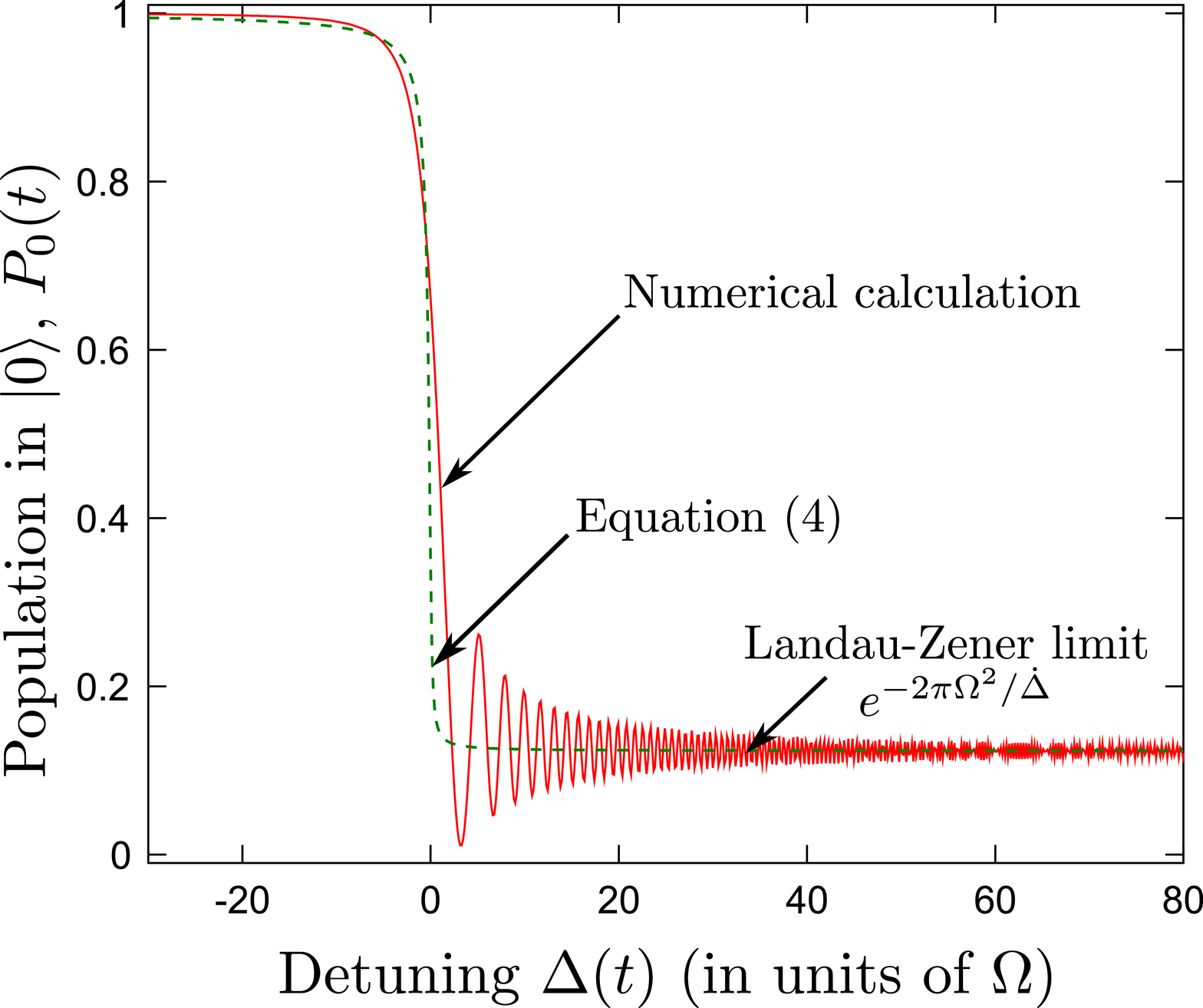}
\caption{Comparison between the analytical estimate in Equation (\ref{eq:2}) and a numerical integration of the Hamiltonian in Equation (\ref{eq:Hint}). The numerical calculation is for a ramp rate $\dot{\Delta} = 3 \Omega^2$, and does not include any spontaneous emission or other dissipative mechanisms. The oscillations seen in the numerical calculation are residual effects of coherent dynamics that are not captured by the Landau-Zener formula.}
\label{fig:two}
\end{figure}

Let $P_0(t)$ be the probability of being in the $|0\rangle$ state at a time $t$. After a time-segment of duration $\delta t$, the probability $P_0(t + \delta t)$ of remaining in the $|0\rangle$ state at the end depends on the transition rate $\Gamma(t)$ within that segment. We have
\begin{equation}
P_0(t + \delta t) = [ 1 - \Gamma(t) \ \delta t ] \ P_0(t) \approx e^{-\Gamma(t) \ \delta t} P_0(t).
\end{equation}
Therefore the probability $P_0(t_f)$ of remaining in $|0\rangle$ over the entire duration of the process, from an initial time $t_i$ to a final time $t_f$, is 
\begin{eqnarray}
P_0(t_f) & = & \exp\Big(-\int_{t_i}^{t_f} \Gamma(t) \ dt\Big) \nonumber \\
& = & \exp\Bigg(-\Omega^2 \int_{t_i}^{t_f} \frac{\gamma}{\Delta(t)^2 + \gamma^2/4} \ dt \Bigg)\nonumber \\
& = & \exp\Bigg(-\frac{2 \Omega^2}{\dot{\Delta}} \int_{\Delta(t_i)}^{\Delta(t_f)} \frac{\gamma/2}{\Delta^2 + (\gamma/2)^2} \ d\Delta \Bigg) \nonumber \\
& = & \exp \Big\{- \frac{2 \Omega^2}{\dot{\Delta}} \Big[ \mathrm{atan}\Big( \frac{\Delta(t_f)}{\gamma/2} \Big)  - \mathrm{atan}\Big( \frac{\Delta(t_i)}{\gamma/2} \Big) \Big] \Big\} \label{eq:2}.
\end{eqnarray}

The mathematical limits $\Delta(t_i) \ll \gamma/2 \ll \Delta(t_f)$ correspond to the physical case where the detuning is ramped from far below resonance, through the level crossing at $\Delta = 0$, ending up far above resonance. In this limit, Equation (\ref{eq:2}) reduces to the LZF: $P_0(t_f) = \exp(-2 \pi \Omega^2/\dot{\Delta})$. In principle, population transfer from $|1\rangle$ to $|0\rangle$ at a rate given by equation (2) must also be considered. It can be neglected in practice because: (a) with the assumed initial conditions, no considerable population is accumulated in $|1\rangle$ until after the resonance is crossed (see figure \ref{fig:two}), and (b) from that point on, the detuning from resonance only continues to increase, making the population transfer back into $|0\rangle$ increasingly improbable.

Note that the population in $|0\rangle$ is completely transferred over to $|1\rangle$ for an infinitesimally slow ramp through resonance ($\dot{\Delta} \to 0 \Rightarrow P_0(t_f) \to 0$). This is an example of adiabatic passage \cite{BKD08}. For any nonzero ramp speed, the probability of a non-adiabatic transition is given by the Landau-Zener formula.

\section*{Acknowledgments}

I am grateful to David DeMille for many helpful discussions about this approach. This work was supported by the National Science Foundation. Daniel Comparat kindly suggested improvements and pointed out that a qualitatively similar treatment using path integrals can be found in \cite{SS93}.

\section*{References}
\bibliography{lz}
\end{document}